\documentclass{PoS}

\title{
Mirror dark matter explanation 
of the DAMA, CoGeNT and CRESST-II data}

\ShortTitle{Mirror dark matter, DAMA, CoGeNT and CRESST-II experiments}

\author{\speaker{Robert Foot}\\
ARC Centre of Excellence for Particle Physics at the Terascale,\\
School of Physics, University of Melbourne,\\
Victoria 3010 Australia.
        E-mail: \email{rfoot@unimelb.edu.au}}

\abstract{
Dark matter might reside in a hidden sector which contains an unbroken
$U(1)'$ gauge interaction kinetically mixed with standard $U(1)_Y$.
Mirror dark matter provides a well motivated example of such a theory.
We show that the DAMA, CoGeNT and CRESST-II experiments can be simultaneously 
explained within this hidden sector framework.  
An experiment in the Southern Hemisphere is needed to test this explanation via a diurnal
modulation signal.
}

\FullConference{36th International Conference on High Energy Physics,\\
                July 4-11, 2012\\
                Melbourne, Australia}

\begin{document}







There is strong evidence for non-baryonic dark matter 
from a variety of astrophysical and cosmological observations.
Efforts to directly detect dark matter have
achieved some very exciting positive results.  
The DAMA/NaI\cite{dama1} and DAMA/LIBRA\cite{dama2} experiments have
observed an annual modulation in their `single hit' event rate 
consistent with dark matter expectations\cite{dm}.
Low energy excesses in the CoGeNT\cite{cogent,cogent2} and CRESST-II\cite{cresst-II} 
experiments have also been reported.

A specific theory is needed to explain these experiments.
One promising idea is that
dark matter resides in a hidden sector which contains an unbroken
$U(1)'$ gauge interaction kinetically mixed with standard $U(1)_Y$.
That such a theory could provide an explanation of the direct detection experiments
has been discussed  
in the context of mirror dark matter\cite{footold}. [References 
and astrophysical/cosmological discussions can
be found in the reviews\cite{review}].  
Our purpose here is to review and update the most recent work\cite{foot2012} on the experimental 
status of mirror dark matter.

Mirror dark matter features a hidden sector
exactly isomorphic to the ordinary sector. 
That is, fundamental interactions are described by the Lagrangian\cite{flv}:
\begin{eqnarray}
{\cal L} = {\cal L}_{SM} (e, \mu, u, d, A_\mu, ...) +
{\cal L}_{SM} (e', \mu', u', d', A'_\mu, ...) + {\cal L}_{mix} \ .
\end{eqnarray}
If left and right chiral fields are interchanged in the mirror sector, then
the theory exhibits an exact parity symmetry: $x \to -x$. 
The bit ${\cal L}_{mix}$ 
contains possible terms coupling the two sectors together, and includes kinetic 
mixing of the $U(1)_Y$  and $U(1)'_Y$ gauge bosons - a renormalizable
interaction\cite{he}.  
This $U(1)$ kinetic mixing induces
photon-mirror photon kinetic mixing:
\begin{eqnarray}
{\cal L}_{mix} = \frac{\epsilon}{2} F^{\mu \nu} F'_{\mu \nu}
\label{kine}
\end{eqnarray}
where $F_{\mu \nu}$ [$F'_{\mu \nu}$] is the field strength tensor for the photon
[mirror photon]. 
This interaction enables charged mirror sector particles of charge $e$
to couple to
ordinary photons with electric charge $\epsilon e$
\cite{holdom}. 
A mirror nucleus, $A'$, with
atomic number $Z'$ and velocity $v$
can thereby elastically scatter off an ordinary nucleus, $A$, with
atomic number $Z$. This imparts an observable recoil energy, $E_R$, with
\begin{eqnarray}
{d\sigma \over dE_R} = {2\pi \epsilon^2 Z^2 Z'^2 \alpha^2 F^2_A F^2_{A'} \over m_A E_R^2 v^2}
\label{cs}
\end{eqnarray}
where 
$F_A$ [$F_{A'}$] is the form factor 
of the nucleus [mirror nucleus] and natural units are used. 

In this theory, galactic dark matter halos are 
composed of mirror particles. 
These particles form a pressure supported, multi-component plasma
containing $e'$, $H'$, $He'$, $O'$, $Fe'$,...\cite{sph}.  
The temperature of this plasma can be estimated from the
condition of hydrostatic equilibrium:
\begin{eqnarray}
T = \frac{1}{2} \bar m v_{rot}^2 
\end{eqnarray}
where $v_{rot}$ is the galactic rotational velocity and
$\bar m = \sum n_{A'} m_{A'}/\sum n_{A'}$ is the mean mass of the
particles in the halo.
Mirror BBN calculations\cite{paolo2} suggests that 
$\bar m \approx 1.1 $ GeV. 
The halo distribution 
of a mirror nuclei, $A'$, is: 
\begin{eqnarray}
f_{A'}({\textbf{v}},{\textbf{v}}_E) 
 = exp(-E/T) = exp(-\frac{1}{2} m_{A'} {\bf{u}}^2/T) = exp(-{\bf{u}}^2/v_0^2)
\end{eqnarray}
where ${\bf{u}} = {\bf{v}} + {\bf{v}}_E$
[${\bf{v}}$ is the velocity of the halo particles relative to the Earth and ${\bf{v}}_E$ 
is the velocity of the Earth relative to the galactic center].
Clearly 
\begin{eqnarray}
v_0[A'] &=& \sqrt{{2T \over m_{A'}}} 
        = v_{rot} \sqrt{{\bar m \over m_{A'}}}\ . 
\label{v0}
\end{eqnarray} 

The differential rate for $A'$ scattering on a target
nuclei, $A$, is 
\begin{eqnarray}
{dR \over dE_R} = N_T n_{A'} 
\int^{\infty}_{|{\textbf{v}}| > v_{min}}
{d\sigma \over dE_R}
{f_{A'}({\textbf{v}},{\textbf{v}}_E) 
\over k} |{\textbf{v}}| d^3 {\textbf{v}} 
\label{55}
\end{eqnarray}
where the integration limit is, in natural units, 
$\ v_{min} \ = \ \sqrt{ (m_{A} + m_{A'})^2 E_R/2 m_{A} m^2_{A'} }$\ .
In Eq.(\ref{55}), $k=v_0^3 \ \pi^{3/2}$, $N_T$ is the number of target nuclei and 
$n_{A'} = \rho_{dm} \xi_{A'}/m_{A'}$ 
is the number density of the halo $A'$ particles. 
[$\rho_{dm} = 0.3 \  {\rm GeV/cm}^3$ and $\xi_{A'}$ is the halo mass fraction
of species $A'$].
The integral, Eq.(\ref{55}), can be evaluated in terms
of error functions and numerically solved.


Detector resolution effects can be incorporated by convolving the rate
with a Gaussian.
The relevant rates for the DAMA, CoGeNT and CRESST-II experiments   
can then be computed and compared with the data.
Note that the expected predominant $H', \ He'$ halo components are too light to give significant signal contributions
due to exponential kinematic suppression. Only heavier `metal' components can give a signal above the detector
energy thresholds.
We assume for simplicity that the rate in each experiment is dominated by the scattering 
from a single such metal component,
$A'$. 
Of course this is an approximation, however it can 
be a reasonable one given the narrow energy range probed in the
experiments [the signal regions are mainly:
2-4 keVee (DAMA), 0.5-1 keVee (CoGeNT),  12-14 keV (CRESST-II)].
With this assumption we find that
$v_{rot} = 200$ km/s  
is an example where all three experiments have overlapping favored regions 
of parameter space.
In this case
a $\chi^2$ analysis of each experiment leads to the
favored regions of parameter space shown in figure 1. 
Details of the analysis are similar to ref.\cite{foot2012} except
that the most recent CoGeNT data with surface event correction are used\cite{cogent2}. 
\begin{figure}
\includegraphics[angle=270,width=11.6cm]{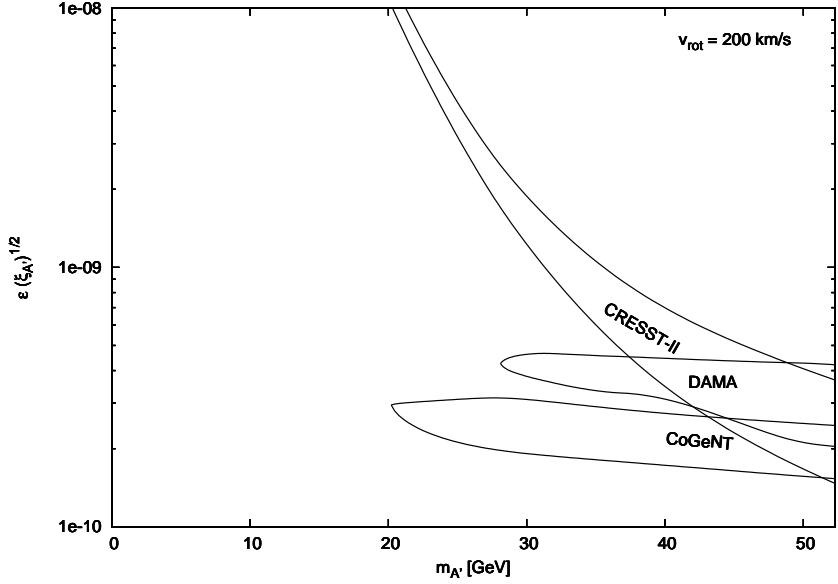}
\caption{
DAMA, CoGeNT and CRESST-II favored regions of 
parameter space in the mirror dark matter model for
$v_{rot} = 200$ km/s.
}
\end{figure}
This figure indicates a substantial region of parameter space where
all three experiments could be explained within this theoretical framework.
An example point, near the combined best fit of the DAMA, CoGeNT and CRESST-II 
data, is:
\begin{eqnarray}
A' &=& {\rm Fe}' \ (m_{Fe'} \simeq 56m_p),\ v_{rot} = 200\ {\rm
km/s}, \ 
\epsilon \sqrt{\xi_{Fe'}} = 2.5\times 10^{-10}\ .
\label{p1}
\end{eqnarray}
The results for this example point are shown in figures 2,3,4. These figures confirm that this type of
dark matter candidate can explain all three experiments simultaneously.
Note that the change in sign of the DAMA annual modulation suggested in figure 2 need not 
happen if there is a lighter and more abundant $A' \sim O'$ component, since the positive
contribution to the annual modulation from $O'$ can outweigh the negative contribution
from $Fe'$\cite{foot2012}.
\begin{figure}
\includegraphics[angle=270,width=12.0cm]{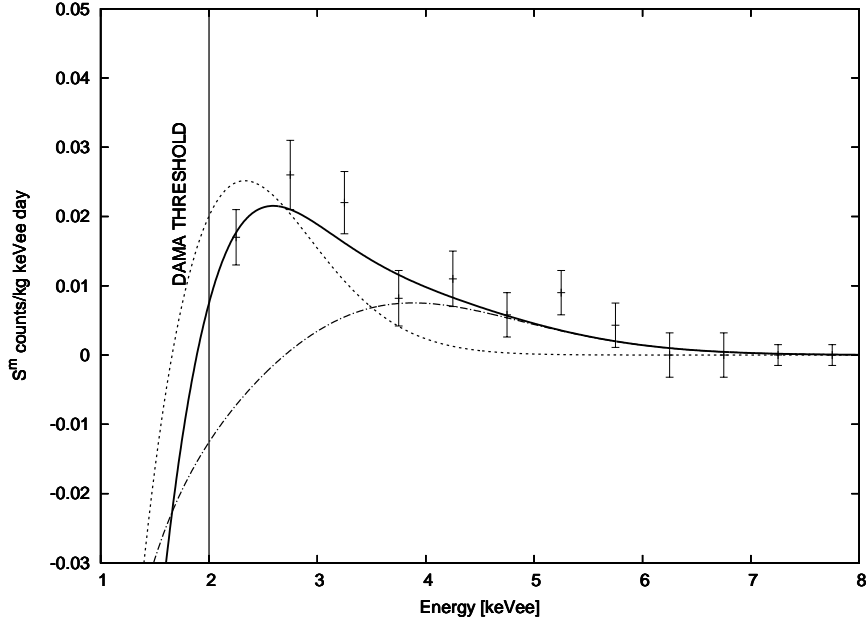}
\caption{
DAMA annual modulation spectrum for mirror dark matter 
with parameter choice, 
Eq.(8) 
(solid line). 
The separate contributions from dark matter scattering off Sodium
(dashed-dotted line) and Iodine (dotted line) 
are shown.   
}
\end{figure}

\begin{figure}
\includegraphics[angle=270,width=12.0cm]{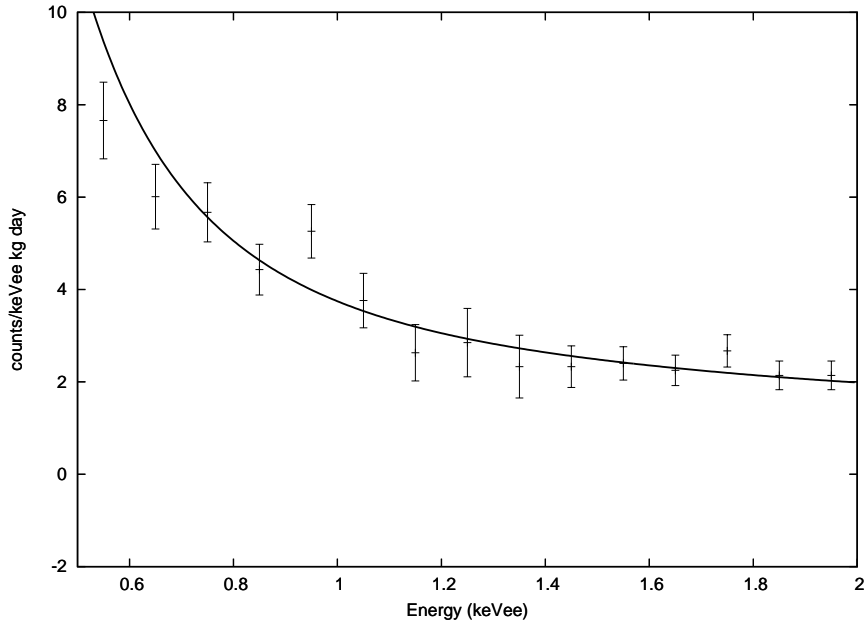}
\caption{
CoGeNT spectrum for mirror dark matter with 
the same parameters as figure 2.
}
\end{figure}
 
\begin{figure}
\includegraphics[angle=270,width=12.0cm]{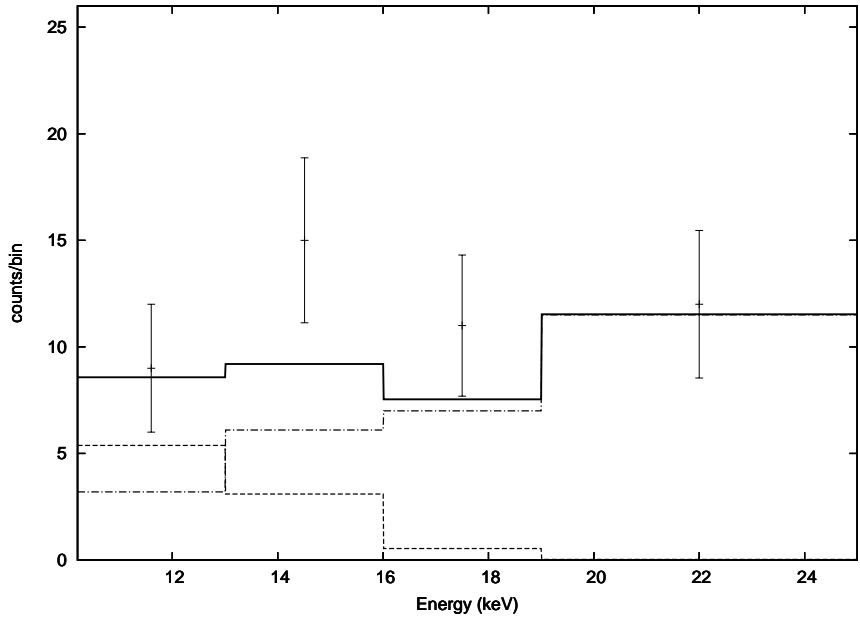}
\caption{
CRESST-II spectrum for mirror dark matter with 
the same parameters as figure 2 (solid line). The signal component (dotted line) and
background component (dashed-dotted line) are also shown.
}
\end{figure}


This mirror dark matter explanation is consistent (although not without some tension) with the null results of the other experiments,
including XENON100 and CDMS, when systematic uncertainties in energy scale are included\cite{foot2012}. 
Future data from DAMA, CoGeNT, CRESST-II and other experiments will
be able to further test and constrain the mirror dark matter framework.
As discussed recently\cite{diurnal},
a particularly striking diurnal modulation signal, shown in figure 5, is predicted 
for a detector located in the Southern Hemisphere. 
Just $\sim 30$ days of operation of the CoGeNT or DAMA detector in say, 
Sierra Grande, Argentina or Bendigo, Australia would
be sufficient to detect the diurnal signal at $5 \sigma$ C.L.

\begin{figure}
\includegraphics[angle=270,width=12.0cm]{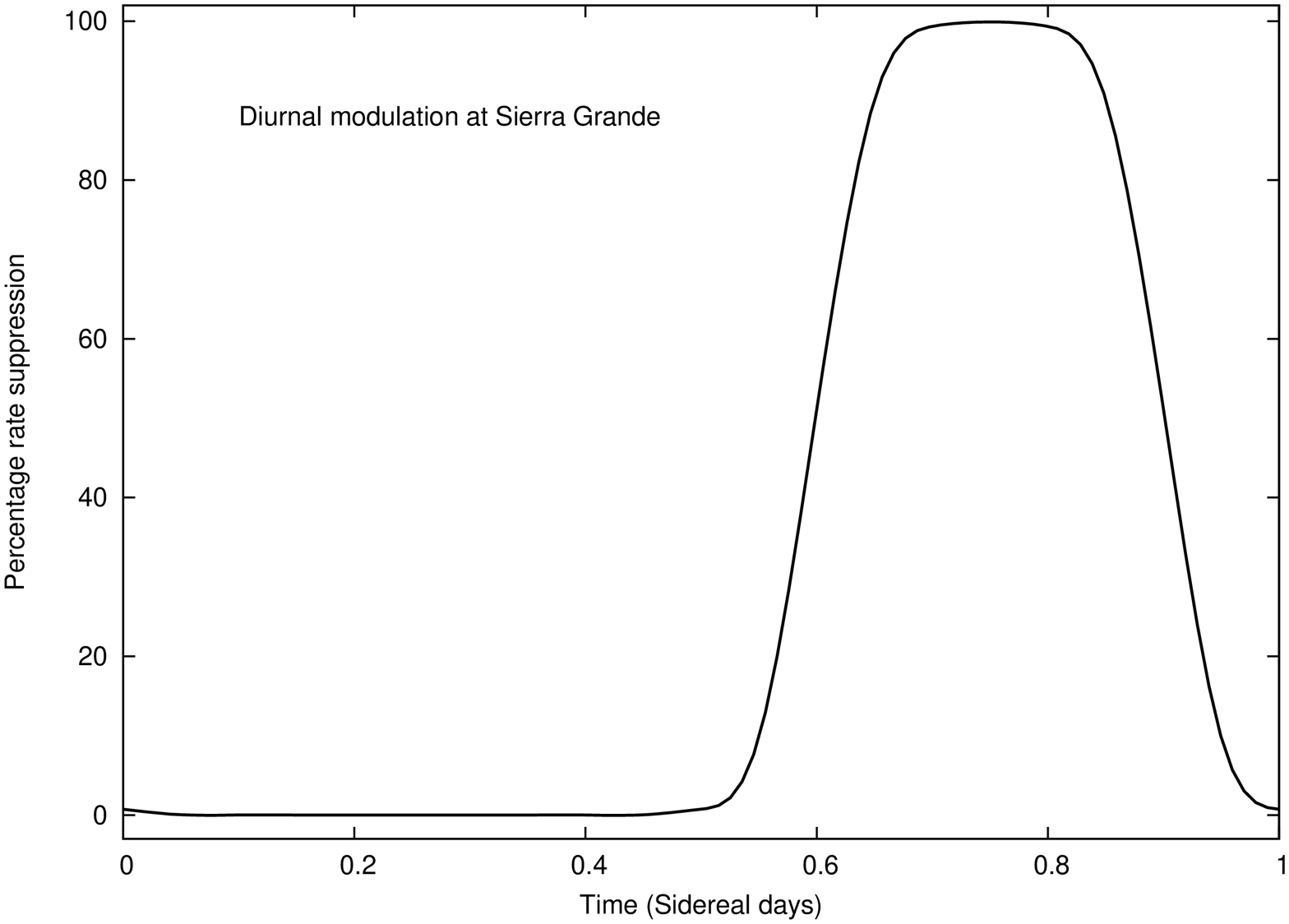}
\caption{
Percentage rate suppression due to the shielding of
dark matter in the Earth's core versus time, for
a detector located at Sierra Grande, Argentina.}
\end{figure}


To conclude,
we have examined the DAMA, CoGeNT and CRESST-II results
in the context of the mirror dark matter framework.
In this scheme
dark matter consists of a spectrum of mirror particles: $e'$, $H'$, $He'$,
$O'$, $Fe'$, ... of known masses.
We have shown that this theory can simultaneously explain 
the data from each experiment by
$A' \sim  Fe'$ interactions
if $\epsilon \sqrt{\xi_{Fe'}} \approx 2\times 10^{-10}$ and 
$v_{rot} \sim 200$ km/s.
Other regions of parameter space, and also, more generic hidden sector
dark matter are also possible.
An experiment in the Southern Hemisphere is needed to test this explanation via a diurnal
modulation signal.

\vskip 0.2cm
\noindent
{\large \bf Acknowledgments}

\vskip 0.1cm
\noindent
This work was supported by the Australian Research Council.


\begin{thebibliography}{99}

\bibitem{dama1}
R. Bernabei {\it et al}. (DAMA Collaboration), 
Riv. Nuovo Cimento. {\bf 26}, 1 (2003); Int. J. Mod.
Phys. E{\bf 13}, 2127 (2004); Phys. Lett. B{\bf 480}, 23 (2000).

\bibitem{dama2}
R. Bernabei {\it et al}. (DAMA Collaboration), 
Eur. Phys. J. C{\bf 67}, 39 (2010);
Eur. Phys. J. C{\bf 56}, 333 (2008). 

\bibitem{dm}
A. K. Drukier {\it et al}, Phys. Rev. D{\bf 33}, 3495
(1986);
K. Freese {\it et al}, Phys. Rev. D{\bf 37}, 3388
(1988).

\bibitem{cogent}
C.~E.~Aalseth {\it et al.}  [CoGeNT Collaboration],
Phys.  Rev. Lett.
{\bf 106}, 131301 (2011); Phys. Rev. Lett. {\bf 107}, 141301 (2011).

\bibitem{cogent2}
C.~E.~Aalseth {\it et al.}  [CoGeNT Collaboration],
arXiv:1208.5737.

\bibitem{cresst-II}
G.~Angloher {\it et al.},
Eur.\ Phys.\ J.\ C {\bf 72}, 1971 (2012).

\bibitem{footold}
R. Foot, Phys. Rev. D{\bf 69}, 036001 (2004); 
Mod. Phys. Lett. A{\bf 19}, 1841 (2004); 
Phys. Rev. D{\bf 78}, 043529 (2008);
Phys. Rev. D{\bf 82}, 095001 (2010);
Phys. Lett. B{\bf 692}, 65 (2010);
Phys. Lett. B {\bf 703}, 7 (2011).

\bibitem{review}
A.~Y.~.Ignatiev and R.~R.~Volkas,
hep-ph/0306120;
R. Foot, Int. J. Mod. Phys. D{\bf 13}, 2161 (2004);
Int. J. Mod. Phys. A{\bf 19} 3807 (2004);
Z.~Berezhiani,
Int.\ J.\ Mod.\ Phys.\ A {\bf 19}, 3775 (2004).
P. Ciarcelluti, Int. J. Mod. Phys. D{\bf 19}, 2151 (2010).

\bibitem{foot2012}
R.~Foot,
Phys.\ Rev.\ D {\bf 86}, 023524 (2012).



\bibitem{flv}
R. Foot, H. Lew and R. R. Volkas, Phys. Lett. B{\bf 272}, 67 (1991);
Mod. Phys. Lett. A{\bf 7}, 2567 (1992).

\bibitem{he}
R. Foot and X-G. He, Phys. Lett. B{\bf 267}, 509 (1991). 

\bibitem{holdom} 
B.~Holdom,
Phys.\ Lett.\ B {\bf 166}, 196 (1986).

\bibitem{sph}
R. Foot and R. R. Volkas,
Phys. Rev. D{\bf 70}, 123508 (2004). 

\bibitem{paolo2}
P. Ciarcelluti and R. Foot, Phys. Lett. B{\bf 690}, 462 (2010).

\bibitem{diurnal}
R.~Foot,
JCAP {\bf 1204}, 014 (2012).

\end{thebibliography}
\end{document}